\documentclass[doublecol]{epl2} 

\title{Mechanical response of active gels}
\shorttitle{Mechanical response of active gels} 

\author{T. B. Liverpool\inst{1} \and M. C. Marchetti\inst{2} \and J-F. Joanny\inst{3} \and J. Prost\inst{4}}
\shortauthor{T. B. Liverpool \etal}

\institute{                    
  \inst{1} Department of Mathematics, University of
Bristol, University Walk, Clifton, Bristol BS8 1TW, U.K.\\
  \inst{2} Physics Department and Syracuse Biomaterials Institute, Syracuse University, Syracuse, NY
13244, USA.\\
\inst{3}Physicochimie Curie (CNRS-UMR168), Institut Curie, Section de Recherche, 26 rue d'Ulm, 75248 Paris Cedex 05, France.\\
\inst{4}E.S.P.C.I., 10 rue Vauquelin, 75231 Paris Cedex 05, France.\\
}
\pacs{nn.mm.xx}{First pacs description}
\pacs{nn.mm.xx}{Second pacs description}
\pacs{nn.mm.xx}{Third pacs description}

\abstract{
We study a  model of an active gel of cross-linked semiflexible filaments with additional active linkers such as myosin II clusters. We show that the coupling of  the elasticity of the semiflexible filaments to the mechanical properties of the motors leads to contractile behavior of the gel,
in qualitative agreement with experimental observations. The motors, however, soften the zero frequency elastic constant of the gel.
When  the collective motor dynamics is incorporated in the model, a stiffening of the network
at high frequencies is obtained. The frequency controlling the crossover between low and high frequency network elasticity is estimated in terms of microscopic properties of motors and filaments, and can be as low as $10^{-3}{\rm Hz}$.
}

\usepackage{bm}  
\usepackage{graphicx}
\usepackage{amssymb}
\usepackage{color}
\usepackage{amscd}
\usepackage{epsfig}
\newcommand{\beq}{\begin{equation}}
\newcommand{\eeq}{\end{equation}}
\newcommand{\beqa}{\begin{eqnarray}}
\newcommand{\eeqa}{\end{eqnarray}}
\newcommand{\bem}{\begin{math}}
\newcommand{\eem}{\end{math}}

\newcommand{\bfr}{{\bf r}}

\newcommand{\bfx}{{\bf x}}

\newcommand{\bfR}{{\bf R}}
\newcommand{\bfrp}{{\bf r}_\perp}
\newcommand{\rp}{ r_\parallel}
\newcommand{\uvec}{\hat{\bf u}}
\newcommand{\uhat}{\hat{u}}

\begin{document}

\maketitle

\section{Introduction}

The mechanical properties of cells control many biological functions, 
including 
the sensing and generation of forces, cell motility and cell division. 
The response of the cell to mechanical stimuli is mediated 
by the cytoskeleton, a network of semiflexible filaments 
(F-actin, microtubules and 
intermediate filaments) linked by a variety of passive and 
active proteins.~\cite{AlbertsBook02,HowardBook00} The cytoskeleton 
is maintained out of equilibrium by chemical reactions that drive 
force generation by motor proteins, as well as by filament treadmilling. 
A variety of recent experiments have measured the remarkable 
rheological properties of this intrinsically nonequilibrium polymer 
network. These include  bulk and microrheology of in vitro stabilized 
networks of cytoskeletal filaments with a controlled concentration 
of various crosslinkers, as well as in vivo whole cell rheology. 

Cross-linked entangled actin networks are viscoelastic solids, 
with a time-dependent mechanical 
response (stress $\sigma$) to deformation (strain $\gamma$). These networks 
have both viscous and elastic responses characterized by 
loss $G''(\omega) \sim \sigma /\dot\gamma$  and storage 
moduli $G' (\omega)\sim \sigma /\gamma$,  respectively. 
For cross-linked gels, the elastic (storage) modulus dominates 
the mechanical response 
 and 
reaches a frequency {\em independent} 
plateau $G_0$ at low frequencies (less than 1Hz). Experimentally $G_0$ 
is found to depend strongly  on cross-link density and can vary 
from 0.1 - 100 Pa~\cite{Gardel04}.
For frequencies above 1Hz, both the storage and loss moduli show 
a high frequency behavior $G',G'' \sim \omega^{3/4}$ characteristic of semiflexible polymer 
dynamics~\cite{IsambertMaggs96}.

Measurements of the mechanical properties of cells yield, however,  
quite different behaviour~\cite{Kaszaa07}. The low frequency ($<10$Hz) 
shear moduli are observed to behave as,
$G',G" \sim G_*(\omega/\omega^*)^\alpha$, with a small exponent 
$\alpha\sim 0.15-0.2$, $G_*\sim 10^2-10^3\;$Pa and $\omega^* 
\sim 1 \;$Hz~\cite{Thoumine97,Fabry01,Wottawah05, Desprat05,Balland06,Stamenovic06}. Significantly, 
the magnitude of $G_*$ is much higher than the typical plateau 
moduli of purified in-vitro actin gels. 
While increasing cross-linker density can significantly enhance the 
elastic modulus ~\cite{Gardel04}, it is surprising that it would 
have such a dramatic effect on the loss modulus. It was recently suggested 
that the remarkable stiffening of the low frequency linear response 
of active gels may be due to the internal stresses generated by the presence 
of active crosslinkers, such as myosin II minifilaments
\cite{Mizuno07,LevineMacKintosh07}. Recent quantitative experiments studying the 
mechanics of in-vitro networks of F-actin,  with passive ($\alpha-$actinin) 
and active (muscle myosin II) cross-linkers, have shown both 
stiffening~\cite{Mizuno07} and contractile behaviour ~\cite{Bendix08} of 
these reconstituted networks.
Interestingly the contractile behaviour has been shown to appear only in 
a narrow concentration range of passive cross-linkers.
 
In this letter we  present a  theoretical description of active 
gels which can explain both 
the contractile behaviour and the intermediate-frequency stiffening of 
these systems. 
The minimal element from which the active gel is constructed is a pair of 
filaments cross-linked by an active cluster of molecular motors. 
This has been a useful starting point for explaining the properties of 
soft active materials 
in both the fluid~\cite{TBLMCM03} and the gel phase~\cite{LevineMacKintosh07}. 
Our work shows that the coupling of the elasticity of 
the semiflexible filaments to the motor dynamics plays a crucial 
role in controlling the rheology of the network.
We find that active clusters lead to 
contraction of the gel which has a more dense ground state than 
a gel with the active crosslinks replaced by passive ones. 
The zero frequency  stiffness of our model active  gel, when perturbed from {\em this ground state}, is lower than that of the corresponding passive gel, in apparent contradiction with experiments.
However, at higher frequencies the collective dynamics of the motors  stiffens the gel as compared  to the passive case.  
This qualititive behaviour is obtained both in 
 the regime of linear chain elasticity and  when taking account of nonlinearities. 

\section{Model}
We consider an ideal semiflexible polymer network with both permanent and
active crosslinkers. The network consists of isotropically oriented 
stiff polymer segments of length shorter than their persistence length 
subjects to rigid constraints due to the permanent
crosslinks. 
The motor clusters act as dynamic cross-linkers which
apply equal and opposite forces to pairs of filaments.
We do not consider the effect of entanglements.

\begin{figure}[h]
\begin{center}
 \resizebox{0.4\textwidth}{!}{%
 \includegraphics{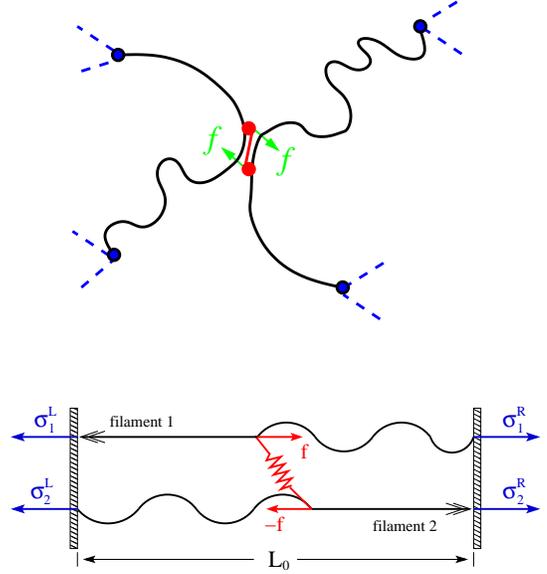}
 }
\caption{Top: A schematic representation of two semiflexible filaments crosslinked by an active  myosin cluster (shown in red) and linked to neighboring filaments (blue dashed lines) by passive linkers (blue dots).  Bottom: The minimal elastic element of our active gel, consisting of two antiparallel filaments crosslinked by a motor cluster. The motor cluster is modeled as a spring of stiffness $k_m$ that exerts equal and opposite forces of magnitude $f$ on the two filaments. The pair of filaments is maintained at a fixed distance $L_0$ by passive crosslinkers. The asymmetric shape of each filament indicates that due to their nonlinear elastic properties the filaments are easier to compress than to stretch.}
\label{fig:schematic}
\end{center}
\end{figure}

We parametrize each filament  by a curve ${\bf R}(s)$, with $0\leq s\leq L$ and 
$L$ the contour length. We consider small deviations from 
a straight configuration
of the polymer segment and decompose deformations of length scales smaller 
than the persistence length $L_p=\kappa/k_BT$, with $\kappa$ 
the bending rigidity,  in transverse and
longitudinal components by writing
${\bf R(}s)=R_{\|}(s)\uvec + \bfr_\perp (s)  \equiv \big[s-r_\parallel(s)\big]
\uvec +\bfrp(s)$,
where $\uvec$ is a unit vector giving the orientation of the segment and 
$\bfr \cdot \uvec =0$~\cite{LiverpoolMaggs01}.
In  a cross-linked gel, the free energy of each filament is given by
\beq\label{wlc_tens}
{F}=\int_0^L
ds\left\{\frac{\kappa}{2}|\partial_s^2\bfR|^2\; - \sigma_0 \uvec \cdot \partial_s \bfR \right\}\;.
\eeq 
The first term is the usual bending energy of a worm-like chain and 
the second one describes
the tension $\sigma_0$ of the filaments due the permanent crosslinkers.
The filament satisfies the boundary conditions
$\bfR(0)=0$, $\bfR(L)=L_0\uvec$ and
$\big[\partial_s^2\bfR(s)\big]_{s=0}=\big[\partial_s^2\bfR(s)\big]_{s=L}=0$,
where $L_0$ is the distance between fixed (passive) crosslinks~\cite{torque}. 
Transverse and longitudinal deviations are coupled by the constraint of
inextensibility,
$\partial_s\rp=\frac12|\partial_s\bfrp|^2+{\cal O}\big(|\partial_s\bfrp|^4\big)$~\cite{LiverpoolMaggs01}.

The  effective longitudinal response function of a filament is evaluated
by averaging over the transverse fluctuations, with the result
\beqa
\left\langle \partial_s r_\| \right\rangle_0 &=& \frac{L}{L_p}\Big\{
{x \coth x -1 \over x^2}\Big\}
 \equiv  {\cal F}(L,\sigma_0,\kappa) \;,
\eeqa
and $x=L\sqrt{\sigma_0\over\kappa}$. The end-to-end length of the filament is
\beq
L_0 = L- \int_0^L ds \left\langle \partial_s r_\| \right\rangle_0  =
 L -\left[\left\langle r_\| (L)\right\rangle_0-\left\langle r_\|
 (0)\right\rangle_0\right] \; .
\eeq
In the limit $\sigma_0\rightarrow 0$ the filament is roughened by 
thermal fluctuations and ${\cal F}
\sim \frac{k_BTL}{6\kappa}\left[1-x^2/15\right]$ for $x\ll1$. 
Conversely, for $x\gg 1$  ${\cal F}\sim k_BT/2\sqrt{\kappa\sigma_0}$ and 
all wrinkles are pulled out by the applied tension so that $L=L_0$.

Now we consider the effect of additional active cross-links  on the mechanical 
properties of the gel. If the gel is kept under constant external tension, the activity 
of the motors changes the end-to-end distance $L_0$.
Conversely, if $L_0$ is changed by applying an external  deformation,
the active cross-links induce an additional  tension  on the 
filaments.
Both types of response may be studied by evaluating the change 
in extension of the filaments upon increasing of the local tension from 
$\sigma_0$ to $\sigma_0+\sigma(s)$. The additional tension may be thought of as arising either 
from motor activity or from externally applied forces. 
To describe this response 
we  define 
an 'elastic' deformation field $u(s)=\left[\langle R_{\|}(s)\rangle
-\langle R_\| (s) \rangle_0\right] $, where $\left\langle R_{\|}(s)\right\rangle$ denotes the effective longitudinal response to the total tension $\sigma_0+\sigma$. In the limit of large $\sigma_0$,
this yields a general relationship between the deformation and the tension,
\beq\label{geneq}
\partial_s u(s) = G\left[\sigma_0,  \sigma(s) \right]\;,
\eeq
with $G={\cal F}(L,\sigma_0,\kappa) -
{\cal F}(L,\sigma_0+\sigma(s) ,\kappa)$. Eq. (\ref{geneq}) describes 
the nonlinear elasticity of a semiflexible filament under tension and is 
the starting point of our analysis.

The top image of Fig.~\ref{fig:schematic} shows a schematic of a motor 
cluster crosslinking two semiflexible filaments, which are in turn bound 
at their ends by permanent crosslinks. 
The motors in the  cluster walk towards the plus end of each filament indicated by 
the double arrows, exerting equal and opposite forces $\pm f$ on the two 
filaments~\cite{TBLMCM03,Mizuno07,LevineMacKintosh07},
resulting in additional tension. 
Such a crosslinked filament pair is the fundamental elastic unit in 
our model of an active gel. 
For an isotropic gel, with uncorrelated orientations of filament pairs 
and motor clusters, the elastic properties of the gel may be obtained 
by suitable angular averages. All the essential physics 
can, however, be obtained from the simplified one-dimensional model depicted in 
the bottom part of Fig.~\ref{fig:schematic}. 
Taking account of the orientation of the filaments only changes 
the numerical prefactors. 
Also we consider only anti-parallel 
filaments as motors do not generate significant tensions on 
pairs of parallel filaments.
Our goal  is to calculate how the end-to-end distance and the 
mechanical response of the  filament pair  are changed by the presence 
of the active crosslink.  The force balance equations for the two 
filaments are
\beqa \sigma_1^R-\sigma_1^L+f=0 \;,  \quad
\sigma_2^R-\sigma_2^L-f=0\; ,
\label{eq:force_fils}
\eeqa
where $\sigma_i^{R,L}$, for $i=1,2$ are the \emph{additional} tensions 
(on top of the "bare" tension $\sigma_0$)
at the ends of each filament
due to active cross-links and/or 
external forces (Fig.~\ref{fig:schematic}).  

\section{Motor clusters} We consider the response of the system 
on time scales long compared  to the relaxation time of 
the {\em longitudinal} modes of the filaments so that we can 
ignore the dynamics of the deformation, $u(s)$. 
We include, however, the finite-frequency response of the motors, 
which is important when 
considering the response of the gel to frequency-dependent  deformations. 
To do this we use the model introduced by 
J\"ulicher and Prost\cite{JulicherProst95}, where a  motor 
cluster actively sliding along a polar filament is described as a 
collection of $N$ motors rigidly attached to a backbone  and moving 
along a polar periodic track.  Under the action of both thermal and 
ATP-driven excitations, each motor in the cluster undergoes transitions
between a strongly bound and a weakly bound state.  
We model our active cross-link as two such motor clusters linked by a 
spring of stiffness $k_m$ that couples the motion along 
the top and bottom filaments.
We denote by $s_i$ the position of the cluster along the $i$-th 
filament and by $u_i=u(s_i)$ the elastic deformation of 
the $i$-th filament at that point and assume that the center of mass 
position  $s_0=(s_1+s_2)/2$ of the motor cluster
does not move. 

The dynamics of the points of attachment of the motor cluster interacting 
with the filaments is described by the force balance equations
\beqa
\zeta_0\partial_t s_1= - \zeta_0\partial_t s_2 =
-k_m(s_1-s_2+u_1-u_2)-f \,, &&\label{s1}
\eeqa
where $\zeta_0$ is a friction. The first term on the right 
hand side of Eq. ~(\ref{s1}) 
represents elastic forces within the motor cluster. 
In the last term, $-f$ is the force exerted by the motors on filament $1$.
In a steady state, where the velocity of the motor cluster vanishes, $f$ equals
the stall force  $f_s$. 
As in Ref.\cite{JulicherProst95}, if the motor clusters are not 
exactly at stall force we expand the force up to first order 
in the motors velocity and introduce an active frequency-dependent friction $\zeta_{act}(\omega)$
on the motors, which can be negative.
For a sake of simplicity, we will assume here that the motor clusters have no spontaneous oscillations, which is the case if the ATP concentration is low enough.

\section{Static response} We first examine the static response of 
a cross-linked filament pair to a change in the end-to-end distance $L_0$. 
We apply a force $2F$ to the filament pair 
and calculate the resultant extension  
of the right-hand side of both filaments $\Delta(F)$. 
A value $\Delta(0)<0$ at zero applied force corresponds to contractile behaviour.

The conformation of each filament is described by a displacement $u_i(s)$, 
with $\partial_su_i=G\left[\sigma_0,\sigma_i(s)\right]$.
These equations must be solved with 
with boundary conditions   $u_i(0)=0$ and
$u_i(L)=\Delta $,  for $i=1,2$. Force balance on the filaments indicate 
that the tension has a jump discontinuity at the point of motor attachment. 
This implies that the derivative  $\partial_su_i(s)$ is also 
piecewise constant, with a jump 
discontinuity of magnitude $f$ at $s_i$. Requiring the displacement 
of each filament $u_i\equiv u_i(s_i)$ at the point of attachment 
to be continuous, we obtain
\beqa\label{conti1}
\frac{u_i}{s_i}=G[\sigma_0,\sigma_0+\sigma_i^L]\;,\\
\label{conti2}\frac{\Delta-u_i}{L-s_i}=G[\sigma_0,\sigma_0+\sigma_i^R]\;,
\eeqa
for $i=1,2$.
Eliminating $u_1$ and $u_2$ from Eqs.~(\ref{conti1}) and (\ref{conti2}), and from the stall condition, 
%
$f_s=f=-k_m(\Delta s+u_1-u_2)$,
%
we obtain
\beqa
&&\label{Delta1}s_1G(\sigma_1^L)+(L-s_1)G(\sigma_1^R)=\Delta\;,\\
&&\label{Delta2}s_2G(\sigma_2^L)+(L-s_2)G(\sigma_2^R)=\Delta\;,\\
&&\label{fm}(s_1-s_2)+s_1G(\sigma_1^L)-s_2G(\sigma_2^L)=-\frac{f}{k_m}\;,\eeqa
where $G(\sigma)\equiv G[\sigma_0,\sigma_0+\sigma]$. From the two force balance equations, Eqs.~(\ref{eq:force_fils}), it is evident that only two of the four tensions $\sigma_i^{R,L}$ are independent. It is convenient to eliminate  two  of the unknowns by introducing new forces 
 $ F _i= \frac12\left(\sigma_i^R + \sigma_i^L\right)$, so that the force balance equations are automatically satisfied. 
Letting $s_{1,2}=s_0\pm\Delta s/2$, $F=(F_1+F_2)/2$ and $\delta F=F_1-F_2$,  Eqs.~(\ref{Delta1}-\ref{fm}) yield a set of three coupled equations in three unknown $\Delta s$, $\delta F$ and  either $\Delta$ of $F$. The equations can be solved to obtain either the displacement $\Delta$ as a function of the total force,   $\Delta(F)$, or $F(\Delta)$. The solution will depend parametrically on the center of mass position of the motor cluster, $s_0$.

We solve Eqs.~(\ref{Delta1}), (\ref{Delta2}) and (\ref{fm}) taking into account the nonlinear elasticity of the filaments, in the perturbative limit where all the motor induced forces are small compared to the bare tension $\sigma_0$  and solve the equations perturbatively in $f$. 
If all tensions $\sigma_i^{R,L}$ are small compared to  $\sigma_0$ one can approximate $G[\sigma_0,\sigma]= \sum_n G^n{\sigma^n / n!} \approx G'\sigma + G''\sigma^2/2 + G'''\sigma^3/6 +\ldots $, where $G^n\equiv G^n(\sigma_0)=\Big(\frac{\partial^n G[\sigma_0,\sigma_0+\sigma]}{\partial\sigma^n}\Big)_{\sigma=0}$, with 
$G'=\big({\partial G[\sigma_0,\sigma]}/{\partial\sigma}\big)_{\sigma=0} $, etc . 
To linear order in the  total force  $F$,
we write
\beq
\Delta( F)=\Delta(0)+\frac{ F}{k_{\mbox{eff}}}+{\cal O}( F^2)\;.
\eeq
The ground state deformation is given by 
\beq
\frac{\Delta(0)}{L}=-\frac{f^2}{2}\Big[\frac{G'}{k_mL}-\phi(1-\phi)\big(G''-2(G')^2\big)\Big]\;,
\eeq
where $\phi=s_0/L$. Since $G'(\sigma_0)>0$ and $G''(\sigma_0)<0$ the ground state deformation is always negative, corresponding to a contractile system.
This result is easily understood if we consider the limit where $G^n=0$ for $n>1$ and the  filaments behave as a linear springs of elastic constant $k_0=(LG')^{-1}$. In this case 
the ground state deformation can be written as $\Delta=fG'\Delta s_0/2$, where 
$\Delta s_0= -\frac{f}{k_m}-2\phi(1-\phi)\frac{f}{k_0}$ is the ground state value of separation of the motor clusters between the two filaments. 
Then $\Delta$ is easily  obtained by equating the change in elastic energy when the filaments are stretched from $L$ to $L+\Delta$, given by $2\big[\frac12 k_0(L+\Delta)^2-\frac12 k_0L^2\big]\simeq 2k_0L\Delta$ to the work $f\Delta s_0$ done by the motor clusters on the filaments.
The effective stiffness of the network is given by
\beqa
\frac{1}{k_{\mbox{eff}}}&=&\frac{1}{k_0}+Lf^2\phi(1-\phi)\Big[G'''/2-2G'G''+(G')^3\Big]\nonumber\\
&&-\frac{f^2}{2k_m}\big[G''+(G')^2\big]\;.
\eeqa
%

\begin{figure}[h]
\begin{center}
\resizebox{0.4\textwidth}{!}{%
  \includegraphics{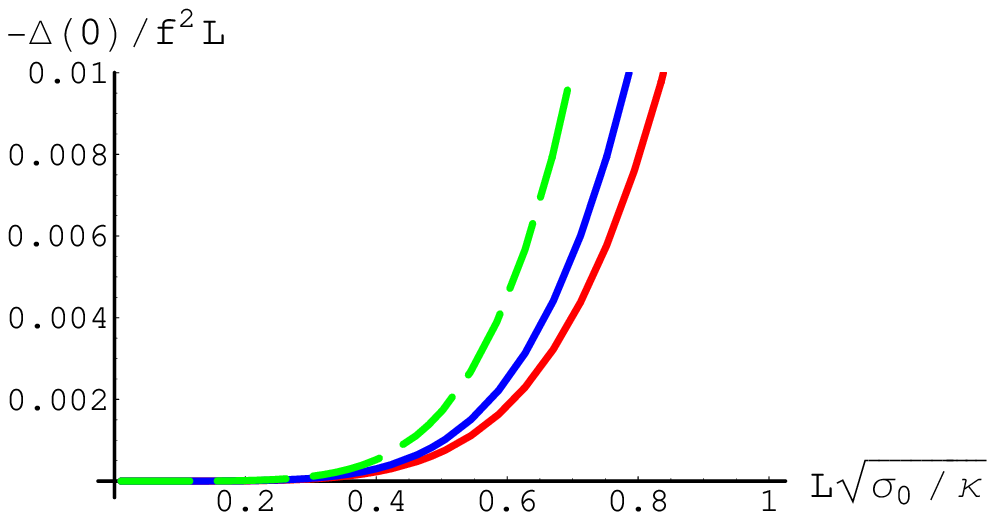}
  }
  \resizebox{0.4\textwidth}{!}{%
   \includegraphics{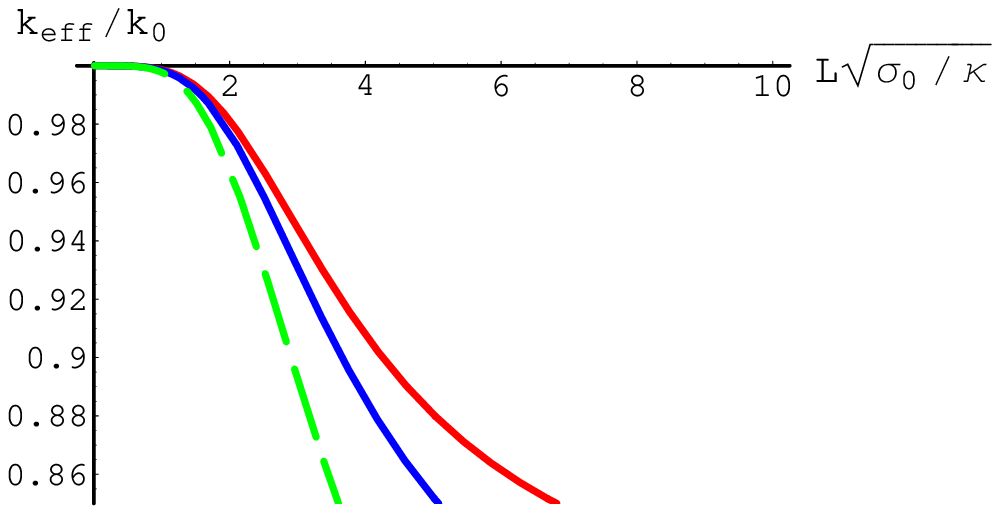}
  }
\caption{Plot of $-\overline{\Delta(0)}/f^2L$  (top) and  zero frequency stiffness $\overline{k_{\mbox{eff}}}$ (bottom) for  $1/k_m=0,{f \over \sigma_0}=0.4$ as a function of $L\sqrt{\sigma_0/\kappa}$ for 3 values of $L_p\sqrt{\sigma_0/\kappa}=$0.2(red), 0.17(blue), 0.13(green).}
\label{fig:Delta-keff1}
\end{center}
\end{figure}

The average ground state deformation and the effective stiffness of the element 
 are shown in Fig. \ref{fig:Delta-keff1} as  functions
of $L$, with $\overline{\Delta(F)}=\int_0^1d\phi\Delta(F)$.
Contractile behaviour is observed for all  $L$ and vanishes as $L \rightarrow 0$, reflecting the higher resistance of short filaments to compression. 
On the other hand, the active crosslinks always
decrease the zero frequency stiffness of the gel which vanishes as $L \rightarrow\infty$.



Our  perturbative analysis captures the qualitative experimental observation  of  
contractile behaviour,  but yields softening of the gel in contradiction with  
experiments. To address this we incorporate in the next section the finite frequency response of the motor clusters.

\section{Finite frequency behaviour}

We now consider the finite frequency behaviour near stall and take into account the 
collective dynamics of the motors 
within an active crosslink cluster.
We consider time-scales long compared to the relaxation time of the filaments, but include the finite-frequency response of the motors.

\begin{figure}[h]
\begin{center}
\resizebox{0.45\textwidth}{!}{%
  \includegraphics{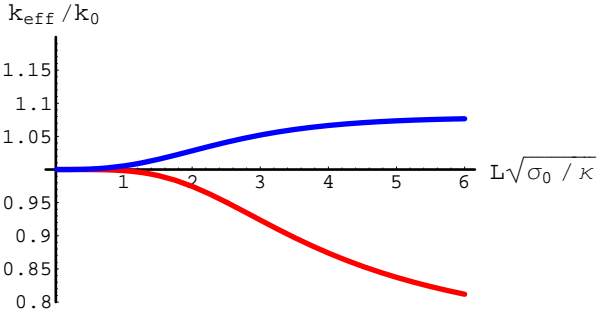} 
  }
\caption{Plot of  $k_{\mbox{eff}}(0)$ (red online) and  $k_{\mbox{eff}}(\infty)$ (blue online) for  $1/k_m=0,{f \over \sigma_0}=0.4$ as a function of $L\sqrt{\sigma_0/\kappa}$ for $L_p\sqrt{\sigma_0/\kappa}=$0.2.}
\label{fig:keffinfty}
\end{center}
\end{figure}

We apply an oscillatory deformation to the end-points of the right side of the filaments, $\Delta=\Delta_0 + \epsilon \Delta_1 (t)$, while keeping the left side fixed. Here $\Delta_0\equiv\Delta(0)$ is the static displacement at zero external force, corresponding to the motors' stall force.
This perturbation will result in motors sliding along the filaments. To discuss the response of the system we work in Fourier space and linearize in all the  deviations
\bem
f \rightarrow f + \delta f_\omega \; , \; 
\Delta s \rightarrow \Delta s + \delta s_\omega \; , \; 
F \rightarrow F_{\omega}  \; , \;  \delta F \rightarrow  \delta F+  \delta F_{\omega}\;, 
\eem
where $f$, $\Delta s$, $F$ and $\delta F$ denote static quantities at stall condition as defined earlier.
From equation (\ref{s1}) the dynamics of the relative displacements of the two motor cluster's heads is given by 
\beqa
i \omega \zeta_0 \delta s_\omega &=& - 2k_m \left\{ 1+\frac12 \left[ G(\sigma^L_1)+G(\sigma_2^L)\right] \right\} \delta s_\omega  - 2 \delta f_\omega \nonumber \\ && - 2 k_m \left[s_1 \delta\sigma_{1\omega}^L G'(\sigma_1^L) - s_2   \delta\sigma_{2\omega}^L G'(\sigma_2^L)\right]  
\eeqa
where $\delta \sigma_{1\omega}^L = F_\omega + \frac12 (\delta F_\omega + \delta f_\omega)$ and 
$\delta \sigma_{2\omega}^L = F_\omega - \frac12 (\delta F_\omega + \delta f_\omega)$.
Solving equations (\ref{conti1},\ref{conti2}) and using the fact that the collective dynamics of the motors induces an active friction~\cite{JulicherProst95} so that $i\omega\zeta_{act} (\omega) \delta s_\omega/2 = \delta f_\omega$, we can obtain a linear relationship between $\delta s_\omega$ and $F_\omega$ and hence 
a frequency dependent correction to $k_{\mbox{eff}}$. 
The expression is complicated but  simplifies in the experimentally relevant regime of 
stiff motors $k_m\rightarrow\infty$ with the result
\beqa
{1 \over k_{\mbox{eff}}(\omega)} = {1 \over k_{\mbox{eff}}(0)} &+&\frac{i\omega L f^2\phi(1-\phi)}
{\omega_c-i\omega}\Big[\frac{(G'')^2}{G'}\nonumber\\
&&-3G'G''+2(G')^3\Big]\;,
\eeqa
where $\omega_c \simeq (\zeta_{act} G'L)^{-1}$. For $\omega \gg \omega_c$ we obtain an enhancement of the effective stiffness of the elastic element:
\beqa\label{keff-infty}
{1 \over k_{\mbox{eff}}(\infty)} \simeq {1 \over k_{\mbox{eff}}(0)} &-& L f^2   \phi(1-\phi)\Big[{(G'')^2 \over G'}\nonumber\\
&&-3G'G''+2(G')^3\Big]\;.
\eeqa
The effective low and high frequency elastic constants are shown in Fig.~\ref{fig:keffinfty}.
Active cross-links always soften the zero frequency stiffness $k_{eff}(0)$ of the elastic element, but at the same time always increase $k_{eff}(\infty)$ relative to the stiffness $k_0$ of a single filament. An estimate of the crossover frequency $\omega_c$ suggests that the stiffening may be relevant at the intermediate frequencies probed in experiments.
\protect
From  ~\cite{JulicherProst95} we estimate $ \zeta_{act} \sim N {1 \over l^2} {\omega_2 (W_2 - W_1)\over (\omega_1 + \omega_2)^2}$, where $N$ is the  cluster size, $W_{1,2}$ are the typical potentials, $\omega_{1,2}$ typical motor conformation  transition frequencies and $l$ the motor's step length.
From the expression for $G$ above we estimate (at $\sigma_0 \simeq 0$), $G'={L^3 \over 90 k_B T L_p^2}$. Finally we obtain 
\bem
\omega_c \sim  \frac{90}{N}{ (\omega_1+\omega_2)^2 \over \omega_2 }   \left({k_BT \over W_2-W_1 }\right) \left( {L_p \over L }\right)^2  \left( {l \over L }\right)^2\, . \, 
\eem
Using  $l \simeq 4 \mbox{nm}$, 
$ W_2 - W_1 \simeq 10 k_B T $,
$L_p \simeq 20 \times 10^3 \mbox{nm} $,  $L\simeq 4\mu{\rm m}$,
 $N \simeq 100  $, and
$ \omega_{1,2} \simeq 100 \mbox{Hz}$,
we obtain
\bem \omega_c \sim 10^{-3} Hz \eem.


We found that both the contractility and the stiffening of 
the active element are proportional to the square of the stall force of 
the motor cluster.  This point and its implication for the macroscopic shear modulus of 
a crosslinked network merit some discussion.

\section{Shear modulus of an active gel}
First we use standard methods to relate the shear modulus of a crosslinked network of noninteracting elastic elements to the  stiffness 
$k_{\mbox{eff}}$  of each elastic  element~\cite{Fredsq,Morse98}.  We describe the active gel 
as a cross-linked semiflexible polymer gel of {\em monomer} density $\rho$,  where each monomer is a sphere of diameter $a \ll L_p$. 
The mesh-size is  then $\xi^2=1/ (\rho a)$~\cite{IsambertMaggs96,Fredsq,Morse98}. Under a uniform shear, a point
$\bfx$ of the network is deformed according to  $\bfx \rightarrow \bfx + {\bf a}(\bfx)$, with 
$\gamma_{ij}=\frac12(\partial_ia_j+\partial_ja_i)$ the applied strain. An elastic segment of orientation $\hat{\bf u}$ and end-to-end distance  $L_0$   undergoes a relative change
 $\delta L_0/L_0 =\gamma_{ij}\uhat_i\uhat_j$.
This deformation will in turn induce a tension $\sigma_s=k_{\mbox{eff}}\delta L_0$ in the elastic unit, where 
$1/k_{\mbox{eff}}$ is the longitudinal response function  of the unit.
The corresponding  contribution to the stress tensor of the gel is 
\bem \sigma_{ij} = \xi^{-2}  \langle \sigma_s\uhat_i \uhat_j\rangle
 \eem,
where  $\langle...\rangle$ denotes an average over the filaments' orientation. For an isotropic filament distribution
$\langle\uhat_i\uhat_j\uhat_k\uhat_l\rangle=(1/15)[\delta_{ij}\delta_{kl}+\delta_{ik}\delta_{jl}
+\delta_{il}\delta_{jk}]$. Defining the shear modulus $E$ of the gel via
$\sigma_{ij} = 2 E \gamma_{ij}$ \cite{Fredsq,Morse98}, for an incompressible gel we obtain
\bem
E= {1 \over 15} \xi^{-2} L_0 k_{\mbox{eff}}.
\eem

The presence of molecular motors acting as active cross-links modifies
both the mean end-to-end length $L_0$ of a filament strand  and the stiffness 
$k_{eff}$ of each elastic element.  As shown earlier, an active crosslink stiffens $k_{eff}$ at high frequencies ($\omega>\omega_c$), corresponding to  time scales shorter  than the relaxation time of the motor clusters, but longer than the relaxation time of a single filament. This correction, given in Eq.~(\ref{keff-infty}), is proportional to the square of the stall force $f$, which in turn is linearly proportional to  
 the ATP activity $\Delta \mu$ (the chemical potential difference between ATP and its hydrolysis products), for small $\Delta\mu$. This will therefore yield an active stiffening of order $(\Delta\mu)^2$ of the shear modulus of the network. 
To estimate the effect of activity on $L_0$  we note that
the network contains a bulk density $n_{\mbox{\tiny pass}}$ of passive crosslinks and a bulk density
$n_{\mbox{\tiny mot}}$ of active crosslinks. Among the latter only a fraction $r=k_{on}/(k_{on}+ k_{off})$ are bound, where $k_{on}$ and $k_{off}$ are the motors binding and 
unbinding rates, respectively.  Assuming only pairwise crosslinks and no dangling ends, the mean strand length between crosslinks is  $L_0=\frac12 \rho a /(r n_{\mbox{\tiny mot}} + n_{\mbox{\tiny pass}})$.
The rates $k_{on}$ and $k_{off}$  depend on  $\Delta\mu$ and 
are finite at chemical equilibrium when $\Delta \mu=0$. 
In general $r$  is expected to depend  linearly on activity for small $\Delta\mu$, i.e., 
$r(\Delta\mu)\approx r_0+r_1\Delta\mu $. This immediately gives
$L _0\simeq L_0^{(0)} + \Delta \mu L_0^{(1)}$,
and will therefore yields active corrections to the shear modulus linear in $\Delta\mu$.

Expanding for small  $\Delta\mu$, the shear modulus of the active gel will have the form $E_{\mbox{\tiny active}}=E_{\mbox{\tiny passive}} +\Delta\mu E^{(1)}+(\Delta \mu) ^2 E^{(2)} +{\cal O}\left((\Delta\mu)^3\right)$. Explicit expressions for the various
contributions can be obtained for instance in the limit of high frequencies  
using the expression (\ref{keff-infty}) obtained earlier for $k_{eff}(\infty)$.
%
The first term, $E_{\mbox{\tiny passive}}$, is the shear modulus of a passive cross-linked semiflexible gel~\cite{Fredsq,Morse98}. The term linear in $\Delta \mu$ comes from the change in the number of crosslinks due to ATP consumption.  The term
quadratic in $\Delta \mu$  has  contributions  from the active forces  (calculated in  this paper),  {\em as well as} smaller contributions from quadratic corrections  to $r$.  It can dominate at intermediate activities where it yields stiffening of the gel at high frequencies.
At small $\Delta\mu$ the correction due to the variation of the binding and unbinding 
rates with activity will dominate. 
The sign of this correction is controlled by the sign of $r_1$ and is
difficult to assess due to two competing effects. 
The unbinding rate $k_{off}$ is 
known experimentally and theoretically to increase with $\Delta \mu$ ~\cite{Mizuno07,parmeggiani}. 
The binding rate $k_{on}$ is also expected to increase with $\Delta \mu$
because once a
motor cluster is bound to one filament, its directed motion along 
the filament allows it to explore a larger region of phase space 
and facilitates the binding to a second filament. 
This can also yield stiffening of the gel at very 
low activity with an elastic modulus increasing linearly with $\Delta \mu$
 if the increase of the binding rate with $\Delta\mu$ dominates the increase in the unbinding rate ($L_0^{(1)} < 0$ above).
Detailed experiments are needed to address this question.

Experiments have found that the addition of active crosslinkers such as myosin II can increase the shear modulus of the network of several order of magnitudes implying that a quantitative comparison requires 
going into the non-perturbative regime.
When the motor induced tension exceeds the bare tension, i.e.,  $f > \sigma_0$, the parts of the filaments under compression will buckle and their response will be governed solely by the bending rigidity, $\kappa$.  
Under such conditions, a complete calculation becomes more difficult. In the "high" frequency regime,
however,  the compressed parts of the filaments contribute negligibly to the force balance
and the modulus is entirely controlled by the tense portion.  A straightforward calculation then yields
$E\simeq2\xi^{-2}(rn_{\mbox{\tiny mot}}/n_{\mbox{\tiny pass}})(\kappa^{1/2}f^{3/2}/k_B T)+ {\cal O}(\frac{\sigma_0}{f},\frac{\kappa}{fL_0^2})$. When $rn_{\mbox{\tiny mot}}\sim n_{\mbox{\tiny pass}}$, the shear modulus of the active gel scales as $(f/\sigma_0)^{3/2}$, in agreement with observations~\cite{Mizuno07,Bendix08}. We stress that all our results only apply if the density of bound motors is larger than a critical density required for a network of tense filaments to percolate trough the gel. For smaller values of $rn_{\mbox{\tiny mot}}$, one expects little effect from the active crosslinks. Conversely, if the average number motor clusters bound to each elastic unit exceeds one, our formulae also breaks down as the portion of filaments between two successive motor clusters do not experience a large tension. These arguments suggest that both motor-induced contractility and stiffening will occur only in a narrow range of density of bound motors,  in qualitative agreement with experiments. 


To summarise, we have studied a simplified microscopic model of a cross-linked active gel and
shown that the nonlinear elasticity and collective dynamics of the motors play an important 
role in the macroscopic mechanical properties of the gel. 
In particular we show that elastic nonlinearities can lead to a gel which is  contractile and 
stiffened by active elements above a characteristic crossover frequency due to the collective 
dynamics of the motors. 

\acknowledgments
We thank Gijsje Koenderink for very useful discussions. 
MCM was supported by  NSF  grants DMR-0305407 and
DMR-0705105 and by the Institut Curie in
Paris through a  Rotschild-Yvette-Mayent sabbatical fellowship. She
thanks the Institut Curie and ESPCI for their hospitality
during the completion of some of this work.
TBL acknowledges the hospitality of the Institut Curie in Paris and the
support of the Royal Society and the EPSRC under grant EP/E065678/1.



\begin{thebibliography}{0}

 
 \bibitem{AlbertsBook02}
\Name{B. Alberts \and A. Johnson \and J. Lewis \and M. Raff \and K. Roberts \and P. Walter}
\Book{Molecular biology of the cell}
\Publ{Garland, New York}
\Year{2002}

\bibitem{HowardBook00}
\Name{J. Howard}
\Book{Mechanics of motor proteins and the cytoskeleton}
\Publ{Sinauer, New York}
\Year{2000}

\bibitem{Gardel04}
\Name{M.L. Gardel \and J.H. Shin \and F.C. MacKintosh \and L. Mahadevan \and
P. Matsudaira \and D.A. Weitz}
\Review{Science}
\Vol{304}
\Year{2004}
\Page{1301}

\bibitem{IsambertMaggs96}
\Name{H. Isambert \and A. C. Maggs}
\Review{Macromolecules}
\Vol{29}
\Year{1996}
\Page{1036}

\bibitem{Fredsq}
\Name{F. Gittes \and F.C. MacKintosh}
\Review{Phys. Rev. E}
\Vol{58}
\Year{1998}
\Page{R1241}

\bibitem{Morse98}
\Name{D.C. Morse}
\Review{Phys. Rev. E}
\Vol{58}
\Year{1998}
\Page{R1237}

\bibitem{Kaszaa07}
\Name{K. E. Kasza \and A. C. Rowata \and J. Liu \and T. E. Angelini \and C. P. Brangwynne \and G. H. Koenderink \and D. A. Weitz}
\Review{Current Opinion in Cell Biology}
\Vol{19}
\Year{2007}
\Page{101}

\bibitem{Thoumine97}
\Name{O. Thoumine \and A. Ott}
\Review{J. Cell Sci.}
\Vol{110}
\Year{1997}
\Page{2109}

\bibitem{Fabry01}
\Name{B. Fabry \and G. N. Maksym \and J. P. Butler \and M. Glogauer \and D. Navajas \and J. J. Fredberg}
\Review{Phys. Rev. Lett.}
\Vol{87}
\Year{2001}
\Page{148102}


\bibitem{Wottawah05}
\Name{F. Wottawah \and S. Schinkinger \and B. Lincoln \and
R. Ananthakrishnan \and M. Romeyke \and J. Guck \and J. K\"as}
\Review{Phys. Rev. Lett.}
\Vol{94}
\Year{2005}
\Page{098103}

\bibitem{Balland06}
\Name{M. Balland \and N. Desprat \and D. Icard \and S. Fereol \and A. Asnacios \and J. Browaeys \and S. Henon \and F. Gallet}
\Review{Phys. Rev. E}
\Vol{72}
\Year{2006}
\Page{021911}

\bibitem{Desprat05}
\Name{N. Desprat \and A. Richert \and J. Simeon \and A. Asnacios }
\Review{Biophysical Journal}
\Vol{88}
\Year{2005}
\Page{2224}

\bibitem{Stamenovic06}
\Name{D. Stamenovi\'c}
\Review{Nature Materials}
\Vol{5}
\Year{2006}
\Page{597}



\bibitem{Mizuno07}
\Name{D. Mizuno \and C. Tardin \and C. F. Schmidt \and F. C. MacKintosh}
\Review{Science}
\Vol{315}
\Year{2007}
\Page{370}

\bibitem{LevineMacKintosh07}
\Name{F.C. MacKintosh \and A.J. Levine}
\Review{Phys. Rev. Lett.}
\Vol{100}
\Year{2008}
\Page{18104}

\bibitem{LiverpoolMaggs01}
\Name{T.B. Liverpool \and A.C. Maggs}
\Review{Macromolecules}
\Vol{34}
\Year{2001}
\Page{6064}

\bibitem{Bendix08}
\Name{P. M. Bendix \and G. H. Koenderink \and D. Cuvelier \and Z. Dogic \and B. Koeleman \and W. M. Brieher \and C. M. Field \and L. Mahadevan \and D. A. Weitz}
\Review{Biophys. J.}
\Vol{107}
\Year{2008}
\Page{117960}


\bibitem{TBLMCM03}
\Name{T. B. Liverpool \and M. C. Marchetti}
\Review{Phys. Rev. Lett.}
\Vol{90}
\Year{2003}
\Page{138102}


\bibitem{torque} 
The condition of zero torque at the filaments' ends is strictly satisfied 
only on average. It suppresses filament buckling under compression.


\bibitem{JulicherProst95}
\Name{F. J\"ulicher \and J. Prost}
\Review{Phys. Rev. Lett.}
\Vol{75}
\Year{1995}
\Page{2618}


  
\bibitem{parmeggiani}
\Name{ A. Parmeggiani \and F. Julicher \and L. Peliti \and J. Prost}
\Review{Europhys. Lett.}
\Vol{56 (4)}
\Year{2001}
\Page{603}


\end{thebibliography}
\end{document}